\documentclass[10pt,conference]{IEEEtran}
\usepackage[letterpaper, top=1in, bottom=1in, left=0.7in, right=0.7in]{geometry}
\RequirePackage[T1]{fontenc}

\usepackage{graphicx} 
\graphicspath{{/.images}}
\usepackage{algorithm}
\usepackage{algpseudocode}
\RequirePackage{mathptmx}
\usepackage{stmaryrd}
\usepackage{bbm}

\usepackage{amsmath}
\usepackage{amssymb}
\usepackage{caption}
\usepackage[table,xcdraw]{xcolor}
\usepackage{color}
\usepackage{tabularray}
\usepackage{cite}
\usepackage{amssymb}

\usepackage{url}
\usepackage{hyperref}

\AtBeginShipout{\AtBeginShipoutUpperLeft{
  \put(\dimexpr\paperwidth-1cm\relax,-1.5cm){\makebox[0pt][r]{\framebox{This paper was presented in the IEEE LCN conference, 2024, Caen, France}}}
}} 

\title{Privacy-Preserving Drone Navigation Through Homomorphic Encryption for Collision Avoidance}

\author{
    \IEEEauthorblockN{Allan Luedeman}
    \IEEEauthorblockA{Dept. of Electrical \& Computer Engineering \\
    Florida International University \\
    Miami, FL, USA \\
    alued004@fiu.edu}
    \and
    \IEEEauthorblockN{Nicholas Baum}
    \IEEEauthorblockA{Dept. of Electrical \& Computer Engineering \\
    Florida International University \\
    Miami, FL, USA \\
    nbaum005@fiu.edu}
    \and
    \IEEEauthorblockN{Andrew Quijano}
    \IEEEauthorblockA{Dept. of Computer Science \& Engineering \\
    New York University \\
    New York, NY \\
    andrew.quijano@nyu.edu}
    \and
    \IEEEauthorblockN{Kemal Akkaya}
    \IEEEauthorblockA{KF School of Computing and Information Sciences \\
    Florida International University \\
    Miami, FL, USA \\
    kakkaya@fiu.edu}
}

\begin{document}

\maketitle

\begin{abstract}
As drones increasingly deliver packages in neighborhoods, concerns about collisions arise. One solution is to share flight paths within a specific zip code, but this compromises business privacy by revealing delivery routes. For example, it could disclose which stores send packages to certain addresses. To avoid exposing path information, we propose using homomorphic encryption-based comparison to compute path intersections. This allows drones to identify potential collisions without revealing path and destination details, allowing them to adjust altitude to avoid crashes. We implemented and tested our approach on resource-limited virtual machines to mimic the computational power of drones. Our results demonstrate that our method is significantly faster and requires less network communication compared to a garbled circuit-based approach. We also provide a security analysis of the approach against potential attacks. 

\end{abstract}

\section{Introduction}
Drone technology is rapidly proliferating in various domains, such as aerial surveillance \cite{drone_surveillance}, package delivery \cite{drone_deliveries}, and search and rescue operations \cite{drone_rescue_2}, \cite{drone_rescue}, with many industries seeing significant advancements. However, data privacy concerns arise as drones are responsible for more services, with one particular instance being the usage of delivery drones. Delivery drones are cost-effective because companies can reduce the costs associated with delivery drivers, fuel, and vehicle maintenance. Drones can also bypass ground traffic and take more direct routes, significantly reducing delivery times compared to traditional delivery methods. Finally, drones can operate outside typical delivery hours, offering faster and more flexible delivery options for consumers.

However, as more delivery drones are being used, collision avoidance becomes a serious concern. Traditional collision avoidance mechanisms often involve sharing detailed flight paths between drones, raising significant privacy concerns. This information may include specific home addresses, shopping patterns, and other data related to individual privacy. If the data is compromised, criminals might target specific homes or businesses for theft or other malicious activities, as they know when valuable packages are expected to arrive. Also, it is important to keep this information private between drones owned by competitors, as leaked delivery data could be used to gain insight into a company's supply chain, customer base, and operational practices, leading to potential business espionage.

To address the need for privacy, various approaches are used, such as homomorphic encryption, secure multiparty computation (MPC) \cite{kemal_paper}, or garbled circuits \cite{multi_robot_paper}. Homomorphic encryption techniques enable parties to perform computations on encrypted data without revealing the underlying information. A garbled circuit is a cryptographic protocol that is used to securely evaluate a function over encrypted inputs without revealing it to the parties involved. It involves converting a Boolean circuit into an encrypted form where the input, output, and internal wires are encoded in such a way that only the intended computation can be performed without leaking any additional information. Li et al. \cite{multi_robot_paper} used garbled circuits to compute whether a collision will occur with two robots. If a collision is detected, both entities would randomize their paths to avoid the collision. The MPC based approach requires a shared grid in which all participating drones must operate, and the processing time escalates with increased size or resolution of the grid, creating scalability issues. Homomorphic encryption approaches, on the other hand, can be customized to the needs of the application without needing extra pre-computations. Depending on the arithmetic operations required, the most efficient homomorphic implementations can be chosen. 

In this paper, using homomorphic encryption, we present a privacy-preserving drone collision avoidance protocol based on the line intersection algorithm \cite{algorithmbook}, which is faster than existing collision avoidance protocols \cite{kemal_paper}\cite{multi_robot_paper} and uses less network bandwidth. 
Our protocol assumes that two drones that have different owners are about to collide. To avoid over-complicating paths, we assume all paths are in a 2-dimensional plane. We utilize an intersection algorithm that operates on encrypted paths using homomorphic encryption, enabling drones to detect potential collisions without revealing the full path of the other drone. The drones would start this protocol once they are within a range that is reasonable enough to cause a collision. If a collision is going to occur, a drone will temporarily change its altitude (that is, modifying the Z coordinate) and return to its default altitude once its path no longer overlaps with the other drone. We are the first to offer the encrypted version of the path comparison algorithm without relying on any pre-computations. 

Our approach has other advantages compared to related work based on our protocol. First, it is an efficient collision avoidance solution that avoids re-computing a path that could cause delays \cite{multi_robot_paper}. Indeed, evaluation results confirm that we can gain about 30\% latency reduction. Also, unlike \cite{kemal_paper}, since we assume we are on a 2-dimensional plane, we can use GPS coordinates as our path, which has the advantage of not requiring any pre-computed matrices for our protocol to work.

The remainder of the paper is organized as follows. In the next section, we provide background information in both homomorphic encryption algorithms and an intersection decision algorithm. Then, Section III discusses related work in the domain of secure route computation. In Section IV, we establish our assumptions, system design, and threat model. Section V discusses our experiment results. Finally, Section VI concludes the paper.

\section{Related Work}

Privacy-preserving collision avoidance has started to receive some recent attention in various contexts. 
For example, Li et al. \cite{multi_robot_paper} developed a privacy-preserving multi-robot planning protocol that would ensure two robots on a floor would not collide. 
A garbled circuit is a function represented by a Boolean circuit with logic gates. Using a two-party MPC approach, one party is assigned the role of \textit{garbler} which generates the garbled circuit, while the other party is called the \textit{evaluator}. The inputs of each party are then encrypted, and the \textit{evaluator} processes the garbled gates with the garbled (encrypted) inputs to compute the garbled output. The \textit{evaluator} can then map this garbled output back to the actual output values using the mapping provided by the \textit{garbler} \cite{garbled}. In Li et al.'s implementation, the output is a Boolean decision on whether an intersection is present somewhere in the complete path. It does not provide exact intersection locations. 

Sciancalepore et al. \cite{george_paper} introduce Privacy-Preserving Trajectory Matching (PPTM), a protocol that enables UAVs to discover potential spatial and temporal collisions without sharing sensitive location and timestamp data with untrusted entities. PPTM employs a tree-based algorithm called Incremental Capsule Matching and integrates a lightweight privacy-preserving proximity testing solution for private comparisons. This approach improves path planning efficiency, reduces delivery time, and minimizes energy consumption for UAVs. However, it does not always guarantee accurate solutions. It trades accuracy and performance. 

Finally, Desai et al. \cite{kemal_paper} propose a secure MPC for trajectory planning among drones, ensuring collision detection without disclosing sensitive information. It uses matrix representation and matrix addition as an approach to compare trajectories, thus reducing computational complexity and improving performance. The matrix representation involves considering each path to take place on a discrete grid of the region. Each drone's path is represented by a series of continuous cells of value one, where anyplace the drone will not pass through is given a value zero. The matrix must be consistent between the drones, and so must cover the entire area of possible flight, and cannot handle traffic that is not familiar with the local matrix, such as a drone from another area.

Compared to the approach of Li et al. \cite{multi_robot_paper}, our approach is faster and lighter on network traffic (as will be shown in Section \ref{sec:performance}), as well as allowing for planning by more than two drones by allowing for deterministic path changes in the event of an intersection rather than random movement. Then, Sciancalepore et al. \cite{george_paper} the most lightweight implementation of their protocol does have false positives and may not be trustworthy in applications where safety is critical such as drone delivery or military surveillance. Finally, compared to Desai et al. \cite{kemal_paper}, our approach eliminates the need for a discrete matrix of paths possibilities, instead taking GPS coordinates as input. This allows for more detailed planning and also bases computation time solely on the complexity of the path, and not on the size of the delivery area being considered. In addition, a drone coming into an area from a far distance could utilize our method because it does not need to have a common matrix of possible locations.

\section{Background}

\subsection{Homomorphic Encryption}
An Additively Homomorphic Encryption scheme such as Paillier \cite{paillier1999public} or DGK \cite{DGK1}\cite{DGK2}\cite{DGK3} allows a user to combine two ciphertexts to receive the sum of both ciphertexts. Assume there are two messages $\alpha$, $\alpha^{'}$ in the plaintext space $\mathbb{M}$ and we define the encryption of $\alpha$ as $\llbracket \alpha \rrbracket$. We use the `boxplus' operator ($\boxplus$) to denote the addition of two ciphertexts such that upon decryption $\llbracket (\alpha + \alpha^{'}) \rrbracket$ = $\llbracket \alpha \rrbracket$ $\boxplus$ $\llbracket \alpha^{'} \rrbracket$.

In Paillier \cite{paillier1999public}, we define $m$ as the plaintext message, $n$ is the product of primes $p$ and $q$, $r$ is a random number and $g$ is a parameter of the encryption scheme, as shown in Eq.  \ref{equation:encrypt}.
\begin{equation}
\label{equation:encrypt}
\llbracket m \rrbracket = g^m r^n \mod{n^2} 
\end{equation}
If you multiply two Paillier ciphertexts, the output would be an encrypted addition of the plain-text. A Paillier ciphertext that is exponentiated with a plaintext value would return the encrypted product of the ciphertext and plaintext as in Eq.  \ref{equation:operations}.
\begin{equation}
\label{equation:operations}
\begin{gathered}
\llbracket \alpha + \alpha^{'} \rrbracket = (g^{\alpha} r^{n})(g^{\alpha^{'}} r^n) = g^{\alpha + \alpha^{'}} r^{{2n}} \\
\llbracket (\alpha)  (\alpha^{'}) \rrbracket = (g^{\alpha} r^{n})^{\alpha^{'}} = g^{(\alpha)(\alpha^{'})} r^{{(\alpha^{'})(n)}}
\end{gathered}
\end{equation}
However, Paillier does not have an operation where you can obtain the product of two ciphertexts. However, using a two-party protocol \cite{encrypted_multiplication} that takes advantage of the distributive property of multiplication, we can obtain the product of two Paillier \cite{paillier1999public} ciphertexts. In this protocol, we assume that Alice has $\llbracket x \rrbracket$ and $\llbracket y \rrbracket$, and generates two random values $a$ and $b$ to additively blind each encrypted value, respectively. Alice sends $\llbracket(x + a)\rrbracket$ and $\llbracket(y + b)\rrbracket$ to Bob to decrypt, multiply, and return the product. Then Alice could use the following operations to obtain $\llbracket xy \rrbracket$, as shown in Eq.  \ref{equation:encrypted_multiply}.
\begin{equation}
\label{equation:encrypted_multiply}
\begin{gathered}
\llbracket xy \rrbracket = \llbracket (x + a)(y + b) \rrbracket \boxplus \llbracket -bx \rrbracket \boxplus \llbracket -ay \rrbracket \boxplus \llbracket -ab \rrbracket
\end{gathered}
\end{equation}

Finally, another application of homomorphic encryption is to compare it to encrypted numbers. Assume that there are two $t$-bit integers held by party A and party B, which are represented in binary form, $x=\sum_{i=0}^{t-1}x_i2^i$ and $y=\sum_{i=0}^{t-1}y_i2^i$. The objective of parties A and B, respectively, is to obtain the protocol bits $\delta_A$ and $\delta_B$, such that $\delta_A \oplus \delta_B={\mathbbm{1}}\{x \le y\}$.

The secure integer comparison problem was first stated in \cite{Yao}, where two millionaires would like to determine who is richer without revealing the amount of their wealth. The first solution to the problem was introduced in \cite{LinAndTzeng} and various more advanced approaches such as the implementation of the DGK cryptography system and its comparison protocol \cite{DGK1}\cite{DGK2}\cite{DGK3}. The high communication cost of these comparison protocols was addressed \cite{Veugen}\cite{Veugen_Correction} and finally \cite{JoyeAndSalehi_Comparison_Protocol} improved them to be resistant to timing attacks.

\subsection{Encryption Library}
We utilize a Java library implemented by Quijano and Akkaya \cite{AndrewQuijano}\cite{Homomorphic_Encryption}, which implements the encrypted equality check \cite{Encrypted_Equality_Check}, multiplication over two homomorphic ciphertexts \cite{encrypted_multiplication} and uses the Joye and Salehi encrypted integer comparison protocol \cite{JoyeAndSalehi_Comparison_Protocol}. The encrypted integer, multiplication, and equality check can be used for pairs of Paillier \cite{paillier1999public} ciphertexts or DGK ciphertexts \cite{DGK1}\cite{DGK2}\cite{DGK3}. We use these functions to implement a privacy-preserving version of a line intersection algorithm \cite{algorithmbook}.

\subsection{Threat Model}
\label{sec:threat}

We assume Alice and Bob, representing two drones, respectively, to be honest but curious, so we expect both parties to faithfully follow the encrypted intersection algorithm, but will attempt to gather as much information as possible from each other such as path information, destination location, etc. We also assume that there may be external parties that will attempt to obtain the path information of the drones. 


As an attack, we consider that Alice may attempt to replicate Bobs' path even if it is encrypted through intersections by defining her own path as a set of very small line segments that cover an entire area to a resolution that will give her Bob's path information, as well as a close replica of a complete path. 


\section{Proposed Approach}

\subsection{System Model and Scenario Overview}
\label{sec:assumptions}

In our scenario, we assume the two drones that are engaging with each other through our protocol to be owned by different companies, e.g., Amazon and UPS. In this scenario, Alice is a drone about to take off, and Bob is either in flight or also ready for take-off. We consider a specific timeframe where Alice can send a request for paths of any other drones that are or will be flying in a neighborhood along with its speed. This would require communication between drones at longer distances than local area networks (e.g. WiFi-like ranges). This can be achieved by employing protocols such as 5G NB IoT \cite{drone_5g} or LoRa \cite{LoRA}. This neighborhood should be much larger than the area covered during the flight to cover all drones that could come into the flight area during flight time. Bob, which is or will be flying in the area Alice is broadcasting to, responds. Note that we expect these drones to have different starting and delivery locations. Thus, it would be unlikely that the same pair of drones would possibly collide multiple times in an honest path. However, we assume that all drones have a default altitude enforced by regulations.

\subsection{Encrypted Comparison of Paths}

\begin{figure}[htbp]
\vspace{-2mm}
\begin{center}
    \includegraphics[scale=0.30, width=7cm, height=2.3cm]{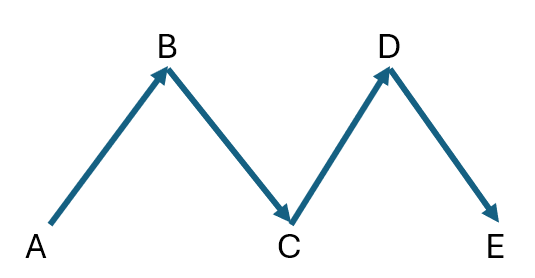}
    \caption{A Path is compromised of an arbitrary number of lines segments, e.g. (A, B), (B, C), (C, D), (D, E)}
    \label{fig:path}
\end{center}
\vspace{-4mm}
\end{figure}

We rely on homomorphic encryption to be able to compare the paths of two drones without exposing any information to each other regarding the path including any intersection point.
In this approach, Bob has a key pair and a route composed of an arbitrary number of line segments. Alice also has a route composed of an arbitrary number of line segments. A sample route along with their line segments is shown in Fig. \ref{fig:path}.

\begin{figure}[htbp]
\begin{center}
    \includegraphics[scale=0.36, width=8cm, height=3.5cm]{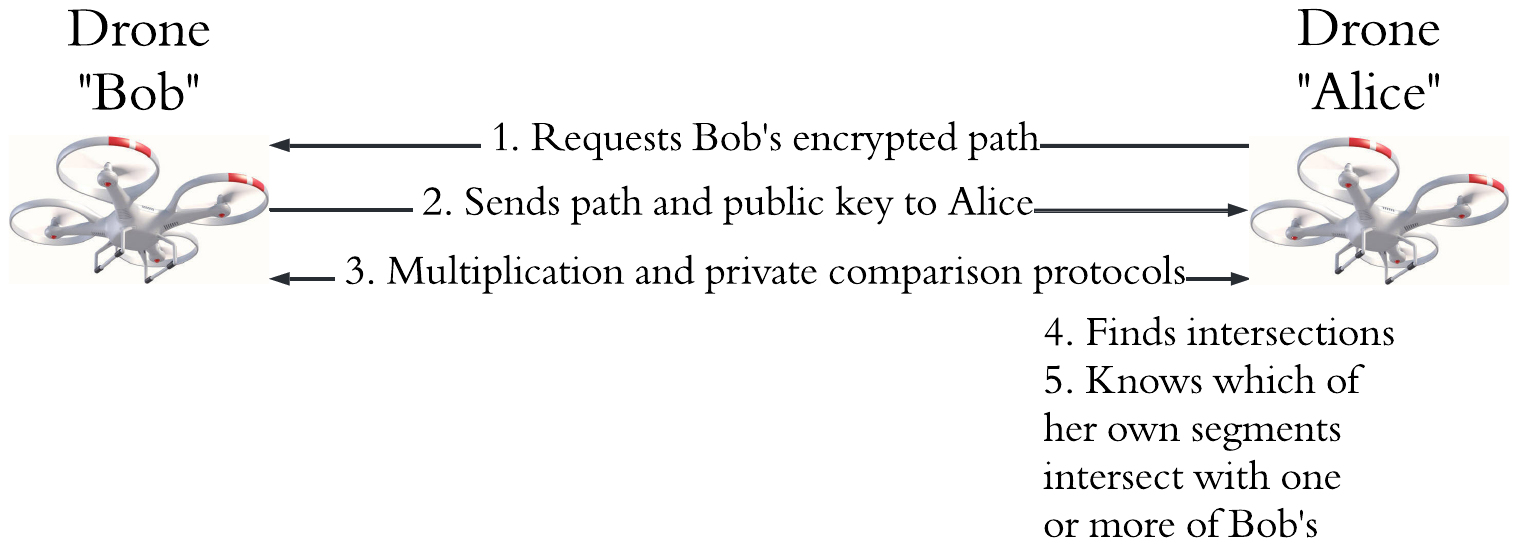}
    \caption{Protocol for two drones to determine which intersections there might be collisions to avoid.}
    \label{fig:system_model}
\end{center}
\vspace{-4mm}
\end{figure}

Both Alice and Bob will follow the steps below to run our approach, as also shown in Fig. \ref{fig:system_model}:

\begin{enumerate}
    \item Bob will encrypt his route using his own public key and await a connection from Alice.
    \item Upon connection to Alice, Bob sends his public key and encrypted route to Alice.
    \item Alice then uses Bobs public key to encrypt her route, and they both complete an encrypted version of the route using Algorithm \ref{algo:intersection}.
\end{enumerate}

Once the algorithm ends, Alice knows which of her line segments on her path will collide with Bob. As Bob's path is encrypted, no other information is available to Alice. Bob does not receive the results of the comparisons and, if he wishes to obtain the same information, the protocol needs to be run again with the roles switched. However, this is not necessary, as it would be assumed that Alice would change altitude to avoid colliding with Bob.



\begin{algorithm}[htb]
\caption{Drone behavior in flight}
\label{alg:flight}
\hspace*{\algorithmicindent} \textbf{Input: Protocol-Initiation-Range} \\
\begin{algorithmic}[1]
\While{\textit{In Flight}}
    \If{\textit{Another drone is in Protocol-Initiation-Range}}
        \State{\textit{Determine which Drone is Alice and Bob}}
        \State{Segments $\leftarrow$ \textit{Encrypted version of Algorithm \ref{algo:intersection}}}
        \For {Segment in Path}
        \If{\textit{Collision occurs at Segment $i$}}
            \State{\textbf{Alice adjusts altitude}}
        \Else
            \State{Return to default altitude}
        \EndIf
        \EndFor
        
    \Else
        \State{\Return{\textbf{continue}}}
\EndIf
\EndWhile
\end{algorithmic}
\end{algorithm}

\subsection{Computing Intersections}
\begin{figure}[htbp]
\vspace{-2mm}
\begin{center}
    \includegraphics[scale=0.4]{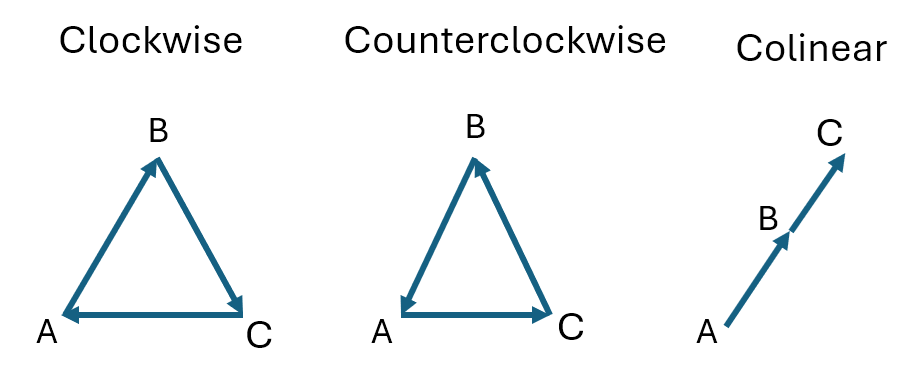}
    \caption{Three cases of orientation for points A, B, and C.}
    \label{fig:orientation}
    \vspace{-2mm}

\end{center}
\vspace{-2mm}
\end{figure}

The line segment intersection algorithm \cite{algorithmbook} (that is, Algorithm \ref{algo:intersection}) we use is based on the concept of orientation of an ordered triplet. There are three possible orientations for any ordered triplet of points: \textit{clockwise}, \textit{counterclockwise}, or \textit{collinear}. To demonstrate this, we used the points (A, B, C), shown in their three possible orientations in Figure \ref{fig:orientation}.

Assume that we have two line segments (A , B) and (C, D) that intersect, there are two cases that would occur.

In the general case, where two line segments intersect, (A, B, C) and (A, B, D) have different orientations and the orientations of (C, D, A) and (C, D, B) are also different. In the other case, if all line segments are collinear, they intersect if their x-projection and y-projections overlap. The pseudocode to find the orientation of an ordered triplet $ \left(A,B,C \right)$ composed of points $ \left(A_x, A_y \right)$, $ \left(B_x, B_y \right)$, and $ \left(C_x, C_y \right)$, where the subscript identifies the plane of projection.

To handle the case that two segments are collinear, we check if any of the four points is located in the other segment.



We can compose a routine that will decide whether two line segments (A, B) and (C, D) intersect. The pseudocode for this is provided in Algorithm \ref{algo:intersection}.

\begin{algorithm}
\caption{Line Segment Intersection Algorithm \cite{algorithmbook}}
\label{algo:intersection}
\begin{algorithmic}[3]
    \State{$\text{O1 = Orientation(A,B,C)}$}
    \State{$\text{O2 = Orientation(A,B,D)}$}
    \State{$\text{O3 = Orientation(C,D,A)}$}
    \State{$\text{O4 = Orientation(C,D,B)}$}
\If {$\text{O1 $\neq$ 02 } \land \text{O3 $\neq$ 04}$}
    \State{\Return{\textbf{True}}}
\EndIf
\If {$\text{O1, O2, O3, O4 are all Collinear}$}
    \If{
    \text{\hskip1.5em A is on segment (C,D) $\lor$} \\
    \text{\hskip1.5em B is on segment (C,D) $\lor$} \\
    \text{\hskip1.5em C is on segment (A,B) $\lor$} \\
    \text{\hskip1.5em D is on segment (A,B) $\lor$}}
    \State{\Return{\textbf{True}}}
    \Else
    \State{\Return{\textbf{False}}}
    \EndIf
\EndIf
\end{algorithmic}
\end{algorithm}

\subsection{Security Considerations}
\label{sec:security}

Since the paths are encrypted, neither Alice nor Bob and no other third party can decrypt any information collected from the wireless broadcasts. 

Now, let us assume Alice will be logging the results of the collision for each line segment in her path with Bob in an attempt to reverse engineer Bob's path. If there is no collision, privacy is preserved. If there is only one collision, Alice would know that one line segment would intersect with Bob. Since there is no time information, Alice would not be able to determine where in the line segment Bob would collide, or Bob's speed or direction.

\begin{figure}[htbp]
    \vspace{-2mm}

\begin{center}
    \includegraphics[scale=0.35, height=3cm, width=7cm]{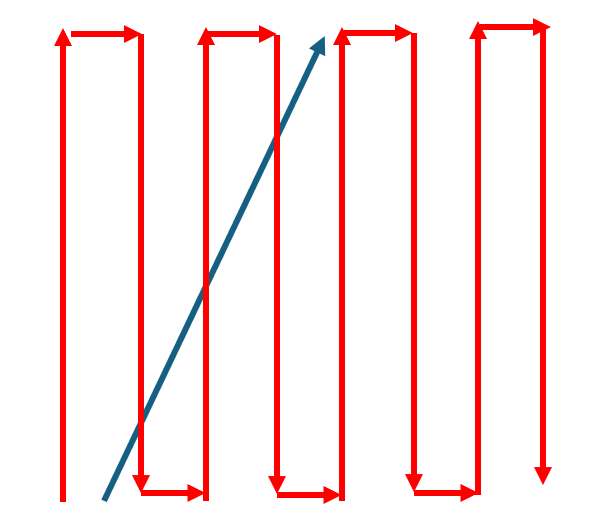}
    \caption{The red path is the path of an Alice drone attempting to brute force, the blue path of a Bob drone.}
    \label{fig:evil}
    \vspace{-2mm}
\end{center}
    \vspace{-2mm}
\end{figure}

During the engagement with Bob, should Alice decide to run the intersection protocol repeatedly against a series of closely spaced parallel lines composed of short segments, as shown in Figure \ref{fig:evil}, she could compose an approximation of Bob's path through the intersections she discovers.

The first mitigation of this attack is that, based on our assumptions in Section \ref{sec:threat}, the drones would have different starting points, so realistically it would still be difficult for Alice to randomly find Bob to attempt this sort of path. This attack also does not factor in that an Alice drone attempting this brute-force approach could just as easily miss fast-moving Bob drones \cite{drone_speed} that came too early or too late within the long parallel line segments. This attack requires Alice and Bob to be within range of each other and to talk to each other for a very long time, which is not possible. Finally, drones also have a very limited range \cite{drone_range}, so if Alice wanted to complete both her delivery and try this task, it would be risky, as the drone increases its risk of running out of energy before returning to its refueling station.

\section{Performance Results}
\label{sec:performance}

\subsection{Experiment Setup}
We implemented our protocol entirely in Java and used a homomorphic encryption library implemented in Java \cite{AndrewQuijano}\cite{Homomorphic_Encryption}. We open-sourced the code we used to implement our collision avoidance algorithm\footnote{https://github.com/adwise-fiu/homomorphic-path-comparison}.

We considered the following metrics to access performance: 
\begin{itemize}
\item \textit{Execution Time}: This is the time for computing if an intersection would occur between two line segments
\item \textit{Network Traffic}: We count the number of bytes that both Alice and Bob write to a socket during the protocols.
\end{itemize} 
Thirty random x-y plane line segments with values from -99 to 99 were generated because \cite{multi_robot_paper} does not provide intersection locations. To fairly compare speed, we used single segments, so both protocols returned the same result: a boolean decision on the existence of an intersection. The experiment was completed on a pair of Debian 12 virtual machines with 2 CPUs and 4 GB of RAM, which were chosen to emulate the resources of a Raspberry Pi 4\footnote{https://www.raspberrypi.com/products/raspberry-pi-4-model-b/}. We calculated the average time and network traffic size for each comparison based on these 30 trials.

\subsection{Experimental Results}

\subsubsection{Comparison with Li et al.}
When comparing our work with Li et al.\cite{multi_robot_paper}, we both used 2048-bit keys. The results in Table \ref{table:results} indicate that our approach is 30\% faster and sends less data over the network compared to Li et al. \cite{multi_robot_paper}. This is because of the efficiency of the comparison operation in DGK. In contrast, MPC-based approaches require a lot of communications and computation to perform comparison operations. The dominating nature of comparison operations in path comparison shows that any time savings should focus on this operation, and use of our approach supports this observation. The gains will even improve with an increase in the number of segments for a given path comparison since only a single segment is focused on average. 

\begin{table}[!htb]
\centering
\caption{Comparing the computation time and bytes sent over a network with \cite{multi_robot_paper}.}
\begin{tabular}{|l|l|l|l|}
\hline
\textbf{Approach}        & \textbf{Time (seconds)}  & \textbf{Network traffic (bytes)}    \\ \hline
Our Approach                        & 4.407s             & 4634                  \\ \hline
Li et al. \cite{multi_robot_paper}   & 6.092s       &  39221                          \\ \hline
\end{tabular}
\label{table:results}
\end{table}
The improvements in speed and network traffic means that our approach would be scalable for multiple pairs of drones to compare their paths within a certain airspace. Our protocol would have the edge if used in networks where bandwidth is limited, such as using LoRa.

\subsubsection{Comparison with Desai et al.}
We also performed an experiment to compare our approach with Desai et al. We used two Raspberry Pi 4s to have a fair comparison (as done in that work). We obtained an average run-time of 16.531 seconds, which is faster than Dessai et al. \cite{kemal_paper} taking 30 seconds for a matrix of size 96 and using a secret share size of 64 bits. 

It is important to understand that the speed of our approach and that of Dessai et al. \cite{kemal_paper} vary based on different parameters. Route complexity does not change the speed of Dessai et al.'s \cite{kemal_paper} approach as it does ours, but the area and resolution considered for route comparisons does. But a matrix size of 96 is very limiting when we consider realistic drone flight areas, and Desai et al. \cite{kemal_paper} computation time scales exponentially with increased matrix size.

\section{Conclusion and Future Work}
In this paper, we implemented a new encrypted intersection algorithm using Homomorphic encryption, which can be used by two drones owned by separate entities to compute where they would intersection. Our algorithm is more flexible than \cite{kemal_paper} as we can use GPS coordinates as part of our line segments, meaning more flexibility on where the protocol could work, as well as not requiring all participating drones to share a finite possible flight area. Furthermore, the results of the experiment showed that our algorithm outperforms the existing approaches. It is faster than both Li et al.'s \cite{multi_robot_paper} collision protocol and Desai et al. MPC protocol and requires less network bandwidth.

The primary privacy concern, a brute force attack designed to reconstruct a flight path out of detected collision points, would take far longer than the flying drone would be in the air, and would fail due to lost connectivity between the drones. The large difference in computation time between an honest path and a brute-force attack also leaves open the possibility of stopping the attack based on assumptions about time and resources used in an honest implementation. 


The aforementioned use of GPS coordinates as input allows for the potential for our approach to be used in-flight in real-time, as opposed to route planning. Using this approach in this way is currently limited by computation time, but as drones increase in computation power and wireless networks increase in bandwidth, it is possible that this approach be used in an in-flight scenario.


\section*{Acknowledgements}
This work was supported by the Griffiss  Institute (Award Number SA10012022020481). The content of this paper is solely the responsibility of the authors and does not necessarily represent the official views of the sponsor. Allan Luedeman and Nicholas Baum would like to thank the Virtual Institute of Cyber Operations and Research (VICOR) for their funding and support. Finally, we thank Alfredo Bayuelo for his assistance in comparing our work with the work of Li et al. \cite{multi_robot_paper}. Andrew Quijano's contribution to this paper was completed outside of his job responsibilities at Amazon. The views expressed in this paper are those of the author and do not necessarily reflect the official policy or position of Amazon.

\bibliographystyle{plain}
\bibliography{references}

\clearpage
\onecolumn
\thispagestyle{empty}  
\begin{center}
    \section*{IEEE Copyright Notice}
\end{center}
\noindent
{\normalsize
© 2024 IEEE. Personal use of this material is permitted. Permission from IEEE must be obtained for all other uses, in any current or future media, including reprinting/republishing this material for advertising or promotional purposes, creating new collective works, for resale or redistribution to servers or lists, or reuse of any copyrighted component of this work in other works.

Published in: \textit{Proceedings of the 2024 IEEE Local Computer Networks (LCN 2024)}.

DOI: \url{https://doi.org/10.1109/LCN60385.2024.10639770}
}
\clearpage
\twocolumn

\end{document}